\documentclass{jfm}
\usepackage[utf8x]{inputenc}
\usepackage[dvips]{color}
\usepackage{graphicx}
\usepackage{graphicx}
\usepackage{natbib}

\ifCUPmtlplainloaded \else
  \checkfont{eurm10}
  \iffontfound
    \IfFileExists{upmath.sty}
      {\typeout{^^JFound AMS Euler Roman fonts on the system,
                   using the 'upmath' package.^^J}%
       \usepackage{upmath}}
      {\typeout{^^JFound AMS Euler Roman fonts on the system, but you
                   dont seem to have the}%
       \typeout{'upmath' package installed. JFM.cls can take advantage
                 of these fonts,^^Jif you use 'upmath' package.^^J}%
      }
  \else
  \fi
\fi

\ifCUPmtlplainloaded \else
  \checkfont{msam10}
  \iffontfound
    \IfFileExists{amssymb.sty}
      {\typeout{^^JFound AMS Symbol fonts on the system, using the
                'amssymb' package.^^J}%
       \usepackage{amssymb}%
         
         \let\geq=\geqslant
      }{}
  \fi
\fi

\ifCUPmtlplainloaded \else
  \IfFileExists{amsbsy.sty}
    {\typeout{^^JFound the 'amsbsy' package on the system, using it.^^J}%
     \usepackage{amsbsy}}
    {}
\fi

\newcommand\Rey{\mbox{\textit{Re}}}  % Reynolds number
\newsavebox{\astrutbox}
\sbox{\astrutbox}{\rule[-5pt]{0pt}{20pt}}

\long\def\beginpgfgraphicnamed#1#2\endpgfgraphicnamed{\includegraphics{#1}}  % content of pgfexternal.tex
\newcommand{\e}{ {\mathrm e} } % exponential

\newcommand{\utot}{{\mathbf u}}

\newcommand{\ubase}{{\mathbf u_0}}
\newcommand{\ubasex}{u_0}

\newcommand{\usuck}{V_S}
\newcommand{\uinf}{U_{\infty}}

\newcommand{\uflucx}{u}
\newcommand{\uflucy}{v}
\newcommand{\uflucz}{w}

\newcommand{\ecf}{{E_{cf}}} % Cross-flow energy
\newcommand{\ecfmin}{\ecf_{min}}
\newcommand{\ecfmax}{\ecf_{max}}
\newcommand{\ecfav}{\langle\ecf\rangle_T}

\newcommand{\burstperiod}{1071}

\title[Edge state for a boundary layer]%
{Edge states for the turbulence transition in the asymptotic suction boundary layer}
\author[T. Kreilos, G. Veble, T.M. Schneider and B. Eckhardt]%
{Tobias Kreilos$^{a,b}$\footnote{tobias.kreilos@physik.uni-marburg.de}, 
Gregor Veble$^{c,d}$\footnote{gregor.veble@ung.si},
Tobias M. Schneider$^{b,e}$\footnote{tobias.schneider@ds.mpg.de}, 
and Bruno Eckhardt$^{a,f}$\footnote{bruno.eckhardt@physik.uni-marburg.de}}
\affiliation{
$^a$Fachbereich Physik, Philipps-Universit\"at Marburg, Renthof 6, D-35032 Marburg, Germany\\ 
$^b$ Max Planck Institute for Dynamics and Self-Organization, Am Fassberg 17, D-37077 Göttingen, Germany\\
$^c$ Pipistrel d.o.o. Ajdovščina, Goriška c. 50a, SI-5270 Ajdovščina, Slovenia\\
$^d$ University of Nova Gorica, Vipavska 13, SI-5000 Nova Gorica, Slovenia\\
$^e$ School of Engineering and Applied Sciences, Harvard University, 29 Oxford Street, Cambridge MA 02138, USA\\
$^f$J.M. Burgerscentrum, Delft University of Technology, Mekelweg 2, 2628 CD Delft, The Netherlands}

\date{\today}

\begin{document}

\maketitle

\begin{abstract}
  We demonstrate the existence of an exact invariant solution to the Navier-Stokes equations for the asymptotic suction boundary layer.
  The identified periodic orbit with a very long period of several thousand advective time units is found
  as a local dynamical attractor embedded in the stability boundary between laminar and turbulent dynamics.
  Its dynamics captures both the interplay of downstream oriented vortex pairs and streaks observed in numerous shear flows as well as the energetic bursting that is characteristic for boundary layers.
  By embedding the flow into a family of flows that interpolates between plane Couette flow and the boundary layer we demonstrate that the periodic orbit emerges in a saddle-node infinite-period (SNIPER) bifurcation of two symmetry-related travelling wave solutions of plane Couette flow.
  Physically, the long period is due to a slow streak instability which leads to a violent breakup of a streak associated with the bursting and the reformation of the streak at a different spanwise location.
  We show that the orbit is structurally stable when varying both the Reynolds number and the domain size.
\end{abstract}


\section{Introduction}
  \label{sec:introduction}
   Recent progress in understanding transitional turbulence and the transition itself is based on the application of concepts from dynamical systems theory. The picture that emerges builds on invariant solutions of the underlying Navier-Stokes equations such as fixed points, travelling waves and periodic orbits. Together with their entangled stable and unstable manifolds these solutions form the backbone of a chaotic saddle that supports the observed turbulent chaotic dynamics: a trajectory approaches the invariant solutions along their stable manifolds and leaves again along the unstable ones; it consists of a chaotic sequence of close visits to the invariant solutions.
  Support for this concept comes in particular from numerical observations in confined geometries, including plane Couette flow \citep{Nagata1990,Schmiegel1997,Kawahara2001,Viswanath2007,Gibson2008a,Halcrow2009}, Taylor Couette flow \citep{Faisst2000} and most prominently pipe flow \citep{Faisst2003,Wedin2004,Eckhardt2008a,Willis2013}, where flow structures very similar to the numerically calculated ones have been observed experimentally by \citet{Hof2004}; see also \citet{Schneider2007a}.
    
  These flows are known to show a transition to turbulence for flow rates
  where the laminar profile is stable against infinitesimal perturbations \citep{Grossmann2000}; finite amplitude perturbations are required to trigger the transition.
  In these systems, the study of dynamically attracting objects within the boundary that separates laminar from turbulent dynamics -- the so called edge states -- has proven very fruitful as these states seem to guide the transition to turbulence \citep{Toh2003,Skufca2006,Schneider2007b,Schneider2008,Duguet2008,Mellibovsky2009,Duguet2010,Kreilos2012}.
  In some confined flow geometries, edge states haven been very well characterized;
  they can be invariant solutions like fixed points in plane Couette flow, travelling waves in symmetry reduced pipe flow or extended plane Couette flow \citep{Duguet2009,Schneider2010a} or be more complex and even chaotic in full pipe-flow \citep{Schneider2007b}.

% The boundary layers that form around bodies in open flows show a somewhat richer dynamics than confined shear flows.
%   In addition to near-wall downstream vortices and streaks, energetic bursting events occur where low-speed fluid is ejected from the wall into the free-stream. These bursts are responsible for the majority of turbulence production in the boundary layer \citep{Kline1967,Robinson1991}. It remains an open question to which extent the
%   invariant solutions can capture both the near-wall vortex structures and the bursting.
  
  The transition to turbulence in boundary layers that form around bodies in open flows shares with the internal flows
  just described the fact that turbulence is observed in parameter ranges where the laminar flow is still stable. 
  Techniques of edge state tracking, initially developed for internal flows, have been adapted and applied to the 
  Blasius boundary layer by \citet{Cherubini2011} and \citet{Duguet2012}. However, limitations on the domain size
  prevented a full characterization of the invariant solutions. More generally, the extension of the concepts developed
  for internal flows to external flows is interesting because of the rich phenomenology seen in external flows.
  For instance, in addition to the near-wall downstream vortices and streaks, which seem similar to the structures
  found in internal flows, energetic bursting events occur where low-speed fluid is ejected from the wall into the 
  free-stream. These bursts are responsible for the majority of turbulence production in the boundary layer \citep{Kline1967,Robinson1991}. The extent to which the
  invariant solutions can capture both the near-wall vortex structures and the bursting in boundary layers remains open;
  see \citep{Kawahara2012} for a recent review.

  As a step towards open flows, we here investigate the turbulence transition in the asymptotic suction boundary layer (ASBL), a boundary layer with a mean normal flow \citep{Fransson2001,Schlichting2004} that shows many features of both internal and external flows. 
  It is ideally suited for an investigation of invariant solutions and edge states in boundary layers, as it has a translationally invariant base flow.  Furthermore, we gain the possibility to connect the system to plane Couette flow by tracking a homotopy parameter, which will allow us to fully characterize and explain the dynamics of the edge state. Finally, the flow connects
  to engineering applications in which suction is a well studied concept for turbulence control, see for example \citet{Joslin1998a}.

   This paper is organized as follows: In \S\ref{sec:system} we describe the flow, its numerical implementation
  and the algorithms used to find edge states. In \S\ref{sec:edgestate} we discuss the
  flow fields for $Re=400$, followed by a discussion of the Reynolds number
  and aspect ratio variations in \S\ref{sec:variation}. Concluding remarks are given in 
  \S\ref{sec:conclusions}.

\section{System, numerical scheme and algorithms}
  \label{sec:system}
  \subsection{The asymptotic suction boundary layer}
\begin{figure}
      \centering
%       \beginpgfgraphicnamed{fig/asbllaminar}
%       \input{pgf/asbllaminar}
%       \endpgfgraphicnamed
      \includegraphics{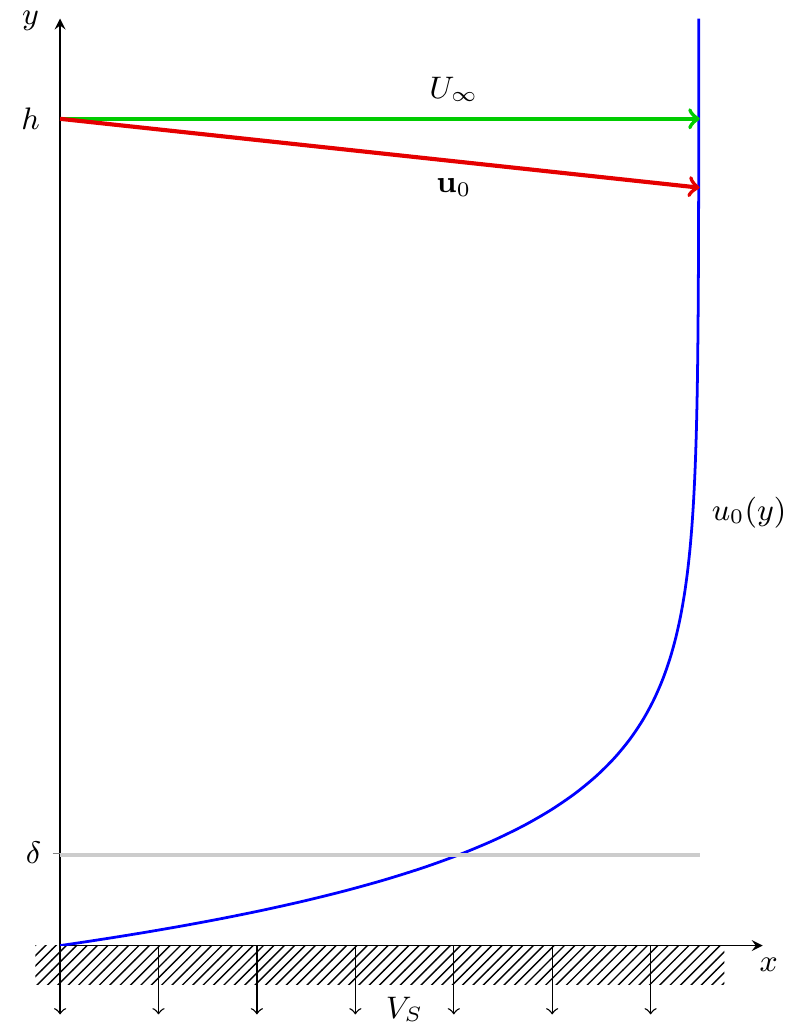}
      \caption{Laminar velocity profile $\ubasex(y)$ for a boundary layer with homogeneous suction $\usuck$ and for $h=10\delta$.}
      \label{fig:asbllam}
    \end{figure}

    The asymptotic suction boundary layer (see figure~\ref{fig:asbllam}) describes a situation where fluid flows with velocity $\uinf$ over a flat plate into which it is sucked with a perpendicular velocity $\usuck$.
    We work in a Cartesian coordinate system, with $x$ the downstream direction; $y$ and $z$ are the wall-normal and spanwise directions, respectively. 
    The equations of motion for an incompressible fluid are the Navier--Stokes equation
    \begin{equation}
      \label{eq:ns}
      (\partial_t + \mathbf{\utot\cdot \nabla})\utot = - \mathbf{\nabla} p + \nu \Delta\utot
    \end{equation}
    together with the continuity equation
    \begin{equation}
      \label{eq:continuity}
      \mathbf{\nabla \cdot \utot} = 0.
    \end{equation}
    
    The no-slip boundary conditions at the wall, the downstream flow and a constant, homogeneous suction $\usuck$ give for the stationary laminar solution to the full Navier--Stokes equation an exponential profile,
    \begin{equation}
%       \label{eq:asbllam}
      \ubase = \left( \uinf(1-\e^{-y/\delta}),\quad -\usuck,\quad 0 \right).
    \end{equation}
    The length $\delta = \nu/\usuck$ equals the displacement thickness, defined by $\int_0^\infty(1-u(y)/U_\infty)\mathrm dy$, and is used as length-scale.
    Velocities are measured in units of $\uinf$ and time in units of $t_u = \delta/\uinf$.    
    The Reynolds number for the ASBL is defined as
    \begin{equation}
      \label{eq:re}
      \Rey = \frac{\uinf\delta}{\nu} = \frac{\uinf}{\usuck}.
    \end{equation}

    The laminar profile (equation \ref{eq:asbllam}) is linearly stable for Reynolds numbers up to a critical value of $\Rey_c = 54382$ \citep{Hocking1975}.
    This is a factor of about $100$ more than for a boundary layer without suction \cite{Schlichting2004}.
    As calculations in this work do not exceed Reynolds numbers beyond $750$, the laminar flow is stable for all our considerations.
    
    For the numerical treatment, the unbounded wall-normal direction is cut off at a height $h$, where a second porous plate that moves with constant velocity $\uinf$ is introduced.
    At both walls we enforce no-slip boundary conditions for the $x$- and $z$-components of the velocity and homogeneous suction for the $y$-component.
    In the spanwise and streamwise directions we employ periodic boundary conditions with periods $L_x$ and $L_z$, respectively.
    The upper wall modifies the downstream velocity in the profile slightly. It now is
    \begin{equation}
      \label{eq:asbllam}
      \ubase = \left( \uinf^*(1-\e^{-y/\delta}),\quad -\usuck,\quad 0 \right),
    \end{equation}
    where $\uinf^* = {\uinf}/({1-\e^{-h/\delta}})$. For $h \rightarrow \infty$ the denominator is $1$ and $\uinf^* = \uinf$.
    The $99\%$ boundary layer thickness slightly depends on $h$; it is $\delta_{99\%} = 4.605\delta$ for $h\rightarrow\infty$ and $\delta_{99\%} = 4.601\delta$ for $h=10$, the box height used in most of this work.
    The laminar velocity profile $\ubasex(y)$ for this height is shown in figure \ref{fig:asbllam}.

%     Experimental realizations of the ASBL showed very good agreement with theoretical predictions, see for example the work of \citet{Fransson2001}.

  \subsection{Numerical aspects}
    \label{sec:channelflow}
    For all DNS (Direct Numerical Simulation) in this work we use the open source channelflow library \citep{channelflow}, developed and maintained by John F. Gibson.
    The code solves the full incompressible Navier--Stokes equations in channel geometries with periodic boundary conditions in the streamwise and spanwise directions and no-slip boundary conditions at the walls.
    It uses spectral discretization with Fourier$\times$Cheby\-shev$\times$Fourier modes in the $x$, $y$ and $z$ directions and finite differencing in time.  %\citep{cfmanual}.
    The code has been modified to allow for the blowing/suction boundary conditions.
    As suggested by the channelflow manual, computations in this work use 2/3-dealiasing in the $x,z$ transforms and calculation of the nonlinear term in the rotational form 
    $(\utot \mathbf{\cdot \nabla}) \utot = (\mathbf{\nabla}\times\utot)\times\utot + \mathbf{\nabla}(\utot \mathbf{\cdot} \utot)/2$.
    For time-stepping a third-order semi-implicit backwards-differentiation method is used. 

    The resolution necessary to resolve all relevant flow structures has been carefully investigated. In a box of size $L_x\times h\times L_z=4\pi\times10\times2\pi$ at Reynolds number $400$ we increased the number of modes until all parameters of two edge states found at two different resolutions were identical to within 1\%. A resolution of $48\times129\times48$ modes was found to be sufficient. The high number of required Chebyshev modes is caused by the need to resolve small fluctuations near the lower wall. 
    For wider and longer boxes the resolution was chosen such that the number of modes per unit length is at least the resolution demanded by the above analysis, i.e.\ $M_x/L_x \geq 48/4\pi$ and $M_z/L_z \geq 48/2\pi$.
    We have examined the influence of the box height on the flow with a similar method and found that a height of $h=10$ is sufficient to calculate edge states. It is not sufficient, though, for simulating fully turbulent flows, which need a much higher domain.
 
    We also found good agreement between our results and independent simulations by Schlatter and Duguet (2010, private communication), see also \citet{Khapko2013}.
    This way we could verify our implementation of both the ASBL modifications in channelflow and the edge state tracking algorithm.

  \subsection{The edge of chaos and the edge state}
    \label{sec:eoc}
    In the asymptotic suction boundary layer turbulent motion is observed while the laminar profile is still linearly stable.
    It is in that sense similar to other parallel shear flows like plane Couette flow \citep{Grossmann2000} or pipe flow \citep{Eckhardt2007} and ideas about the state space structure can be carried over from these systems.    
    There is a basin of attraction around the stable laminar profile in which all initial conditions evolve towards the laminar flow. For sufficiently high Reynolds numbers and sufficiently large perturbations a second region appears in which the flow evolves towards a turbulent state (for our purposes here it does not matter whether this turbulent state is transient or persistent).
    In between those two regions is a laminar-turbulent boundary, the edge of chaos \citep{Skufca2006}, which is characterized by initial conditions that neither decay nor become turbulent.
    The edge of chaos forms an invariant manifold of codimension one that separates the basin of attraction of the laminar state from the turbulent region of state-space.
%     It has one intrinsic unstable direction. States that are inside the edge will remain there forever, but states that are minimally off initially will be exponentially driven away from it -- either decaying to the laminar fixed point or increasing to the turbulent state.

%     In many studies, an attracting set to which edge trajectories are drawn has been found.
    States within the edge of chaos evolve towards a relative attractor, the so-called edge state.
    This may be a fixed point as in plane Couette flow for low Reynolds numbers \citep{Schneider2008}, a periodic orbit as in some parameter regions of plane Poisseuille flow \citep{Toh2003} and some low dimensional models \citep{Skufca2006} or a relative chaotic attractor as in pipe flow \citep{Schneider2007,Schneider2007b}, see also \citet{Vollmer2009, Mellibovsky2009, Duguet2009, Duguet2010}.
%     This attracting set is called the edge state. 
%     It is attracting for states inside the edge but has one intrinsic unstable direction perpendicular to it.
    Its unstable manifold has codimension one, so that the edge of chaos is the stable manifold of the edge state.

\begin{figure}
      \centering
%       \beginpgfgraphicnamed{fig/etalgo4}
%       \input{pgf/etalgo4}
%       \endpgfgraphicnamed
      \includegraphics{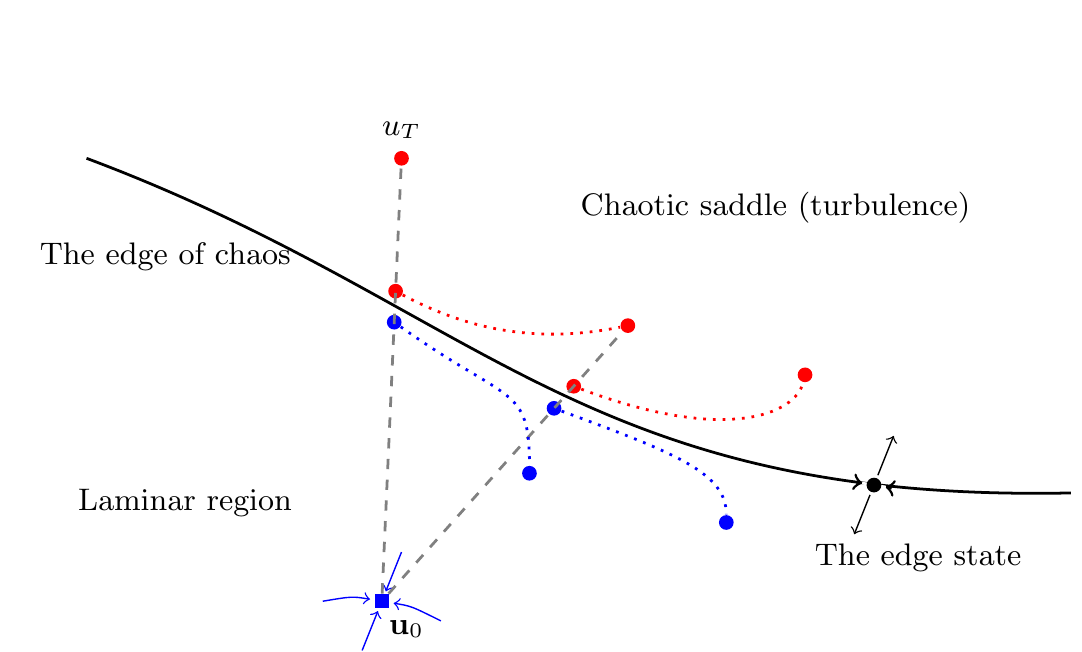}
      \caption{Schematic representation of the edge tracking algorithm. The black line represents the edge of chaos, the laminar fixed point $\ubase$ is at the bottom. A bisection is performed on the line between the laminar and a known turbulent state. The states that bracket the edge are then propagated in time until their separation exceeds a threshold. Then, a refining bisection along the line connecting the new turbulent state and the laminar fixed point is carried out.}
      \label{fig:etalgo}
    \end{figure}

    In order to find the edge of chaos and the edge state we use the technique described in \citet{Schneider2007b}, see also \citet{Toh2003,Skufca2006}.
    For this edge tracking, we classify states that directly decay to the laminar state as lying on the {\it ``low side''} of the edge and states that become turbulent as lying on the {\it ``high side''}. 
    Starting with a turbulent state $u_T$ (on the high side) and the laminar fixed point $u_0$, we use a bisection method to determine two states $u_L$ and $u_H$ on either side of the edge of chaos which are arbitrarily close to it. 
    We can then propagate these states in time, always knowing that the edge lies between them. 
    The two states quickly separate in the edge's unstable direction; after some time, we need a refining bisection to find a new pair of states that are closer together.
    The edge state tracking algorithm consists of a constant repetition of bisection and time integration,
    see figure \ref{fig:etalgo} for a schematic representation.

    As indicated in figure \ref{fig:etalgo}, refining bisections are performed along the line connecting $u_H$ and $u_0$ rather than between $u_H$ and $u_L$, because this method leads to better convergence.
    The reason for this improved performance seems to be connected with the folds in the edge, see \citet{Moehlis2004,Schneider2007b,Vollmer2009}:
    Bisecting between $u_H$ and the laminar fixed point seems to increase the probability of avoiding these folds and of staying near the edge.
    In our simulations we bring the relative $L_2$-distance of $u_L$ and $u_H$ down to $10^{-6}$ and perform a refining bisection once it is larger than $10^{-4}$.

    We classify flow states as lying on the low or high side of the edge by thresholds of the cross-flow energy, i.e.\ the energy in the in-plane velocity components,
    \begin{equation}
      \label{eq:ecf}
      \ecf = \frac 1{L_x L_z \delta}\int_{box} \left(\uflucy^2 + \uflucz^2\right)\mathrm dV,
    \end{equation}
    as a measure of the fluctuations perpendicular to the mainly downstream laminar flow.
    $\ecf$ should be a better indicator for the presence turbulence than just taking the $L_2$-norm of the velocity field since it avoids the dominant contribution from the downstream component and focuses on the perpendicular parts that are necessary to sustain the three-dimensional turbulence.
%     
%     \begin{figure}
%       \centering
%       \includegraphics{fig/et_example}
%       \caption[Trajectories on each side of the edge of chaos]{The energy in the cross-flow velocity components $\ecf$ vs. time for trajectories on each side of the edge of chaos.}
%       \label{fig:example_eoc}
%     \end{figure}
%     
%     Figure \ref{fig:example_eoc} shows the cross-flow energy $\ecf$ for trajectories on each side of the edge of chaos.
    States with $\ecf$ greater than an appropriately chosen threshold become turbulent, while states with $\ecf$ smaller than another threshold decay directly. The use of $\ecf$ as a classification for states proved reliable and well-defined in our simulations, justifying the choice a posteriori. 

\section{The edge state}
  \label{sec:edgestate}
  In this section, we discuss the results of edge state tracking at $\Rey=400$ in a box of size $L_x\times h\times L_z=4\pi\times10\times2\pi$ and report the properties of the edge state. 
  The resolution for this simulation was $48\times129\times48$.
  A large number of edge state trackings for different initial conditions in this box have been performed.
  All initial conditions that converged to some state with regular structure converged to the same state, up to trivial translations, which we hence call the edge state.
  For infinite integration times, we expect all initial conditions to finally approach that state.
%   Some did not converge to any recognizable state (the dynamical properties of edge tracking will be part of a separate study), but all of those who did, converged to the same state, up to trivial translations -- which we hence call the edge state.

  \subsection{Description of the dynamics}
    \label{sec:edgestatedyn}
    \begin{figure}
      \centering
      \includegraphics{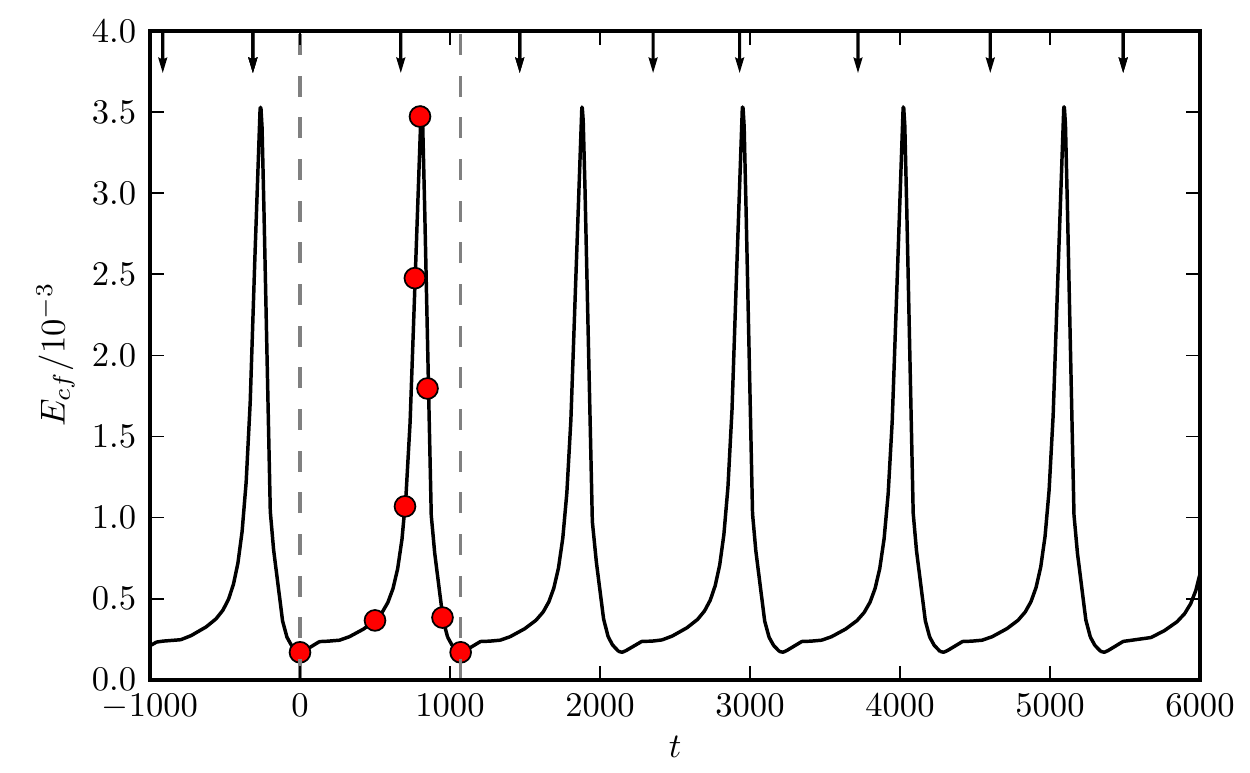}
      \caption{Time evolution of the cross-flow energy of the edge state. The initial transient behaviour is not shown. The arrows at the top indicate the times when a refining bisection was performed, the circles indicate the times where the snapshots in the following figures have been taken. The locations of two minima are indicated with dashed lines; their distance is $1071$ time units.}
      \label{fig:re400:edgestate}
    \end{figure}

    Figure \ref{fig:re400:edgestate} shows the energy in the in-plane velocity components, $\ecf$,  as a function of time. 
%     Only the high field (the field that is on the high side of the edge and becomes turbulent) is shown, the energy of the low field being visually indistinguishable.
    Only the trajectory eventually becoming turbulent is shown, the laminarizing one being visually indistinguishable.
    An outstanding feature of this plot are the periodic strong bursts during which the cross-flow energy $\ecf$ increases almost tenfold. 
    These bursts provide the main starting point for understanding the edge state dynamics.
    The time interval between bursts is $T = \burstperiod$ time units $\delta/\uinf$. 
    While the figure only shows six bursts in approximately 6000 time units, we follow the dynamics for more than 80000 time units to confirm the strict periodicity of the bursts.
%     , we have calculated the trajectory for more than $60 000$ time units, and did not find any changes to the periodicity.}
%     It is very long compared to the expected time scale of the internal flow dynamics -- and also very long for direct numerical simulations. 

\begin{figure}
      \includegraphics{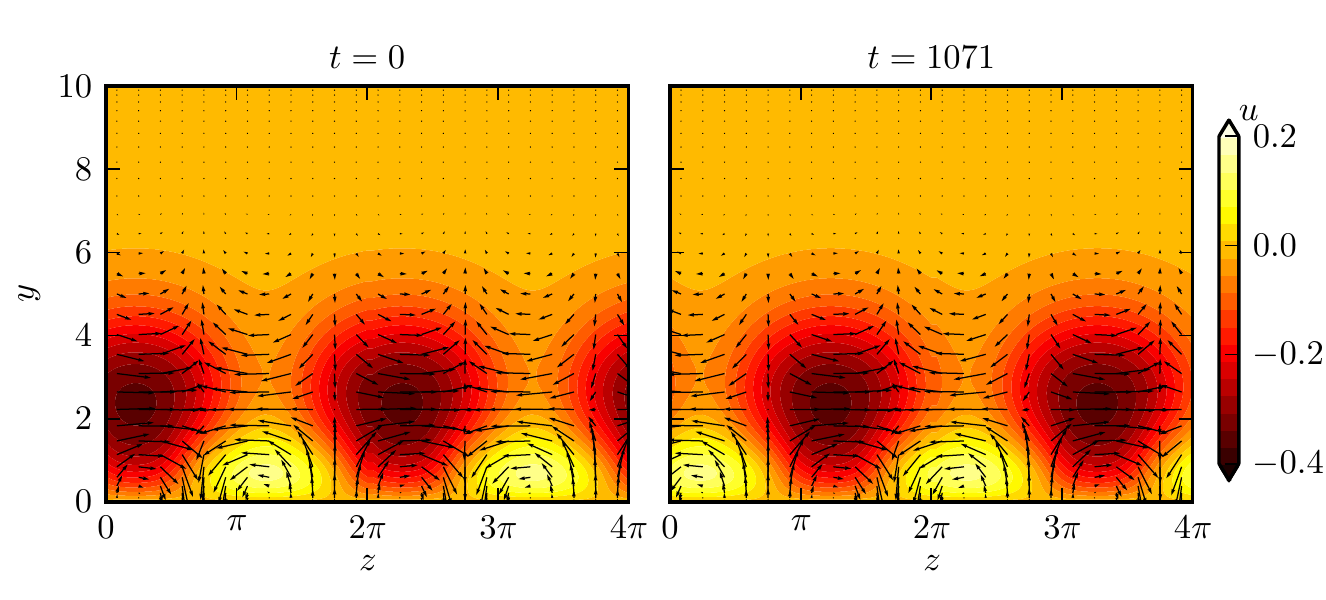}
      \caption{Fluctuating velocity field of the edge state at two successive minima of $E_{cf}$. The velocity field is averaged in the downstream direction, the downstream component is colour coded, in-plane components are indicated with arrows with length proportional to the in-plane vector amplitude. The two images coincide when shifted by $\pi = L_z/2$, indicating a periodic orbit with period $2142$ time units.}
      \label{fig:re400:xaverage}
    \end{figure}

    Figure \ref{fig:re400:xaverage} shows the deviation of the flow from the laminar profile averaged over $x$ at the minimum of $\ecf$ and exactly one burst later. 
    To help visualizing structures across the periodic boundaries in $z$ we show the box twice.
    This representation suggests that after $T=1071$ the flow is the same except for a shift by $\pi$, corresponding to half a box width. After a second burst, the structures are hence back to their original location. 
    Taking the changes in velocity into account, the full period is thus twice the time interval between bursts in $\ecf$.
    To verify that we have an exact invariant solution, we calculate the $L_2$-distances between three fields: the field at t=0, 40 bursts later and 80 bursts later, i.e. after more than 80000 time units.
      The relative distances, $L_2(u_0-u_{40})/L_2(u_0) = 7.86\cdot10^{-6}$, $L_2(u_0-u_{80})/L_2(u_0) = 7.94\cdot10^{-6}$ and $L_2(u_{40}-u_{80})/L_2(u_{40}) = 5.01\cdot10^{-6}$, are very small; the state is indeed an invariant solution to the Naviers-Stokes equations.
    
%     we calculate the $L_2$-distance between $u(0)$ and the field 30 bursts later, i.e.\ after more than 30000 time units. The relative distance, $L_2(u_0-u_1)/L_2(u_0) = 6.7\cdot10^{-05}$, is is very small. }

\begin{figure}
     \centering
     \includegraphics{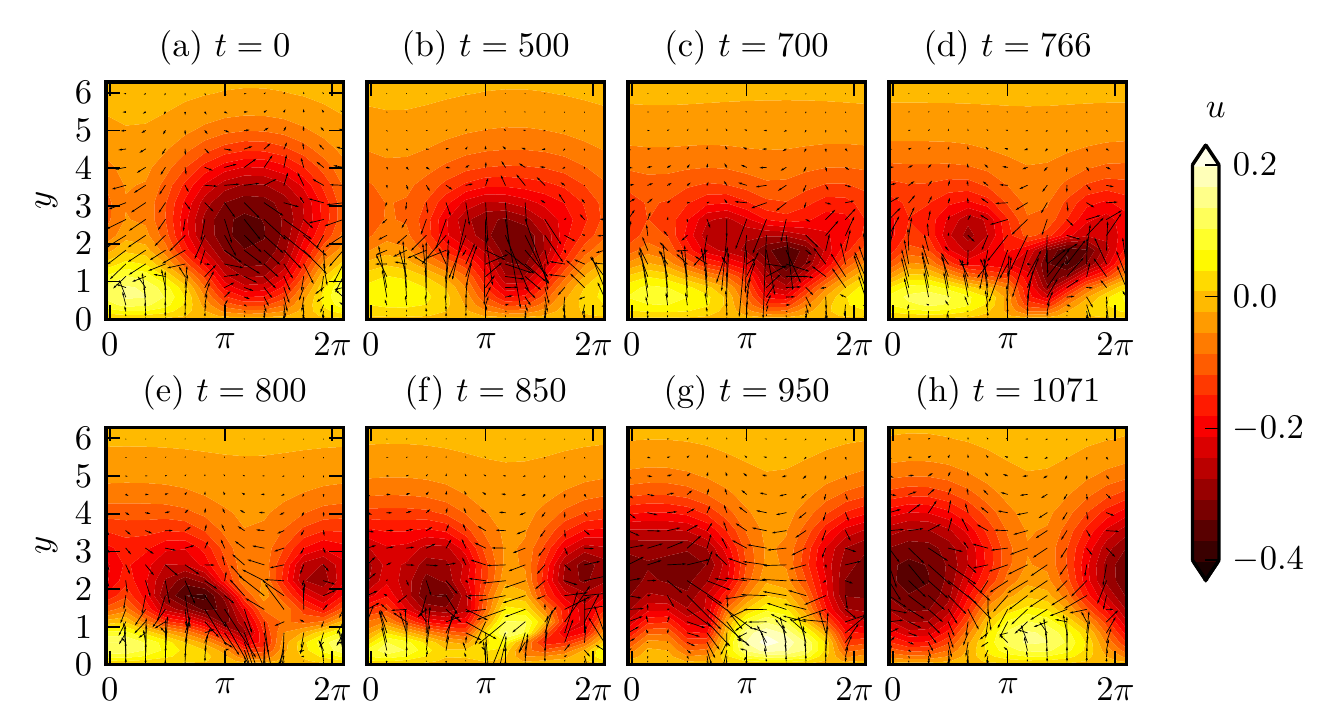}
     \caption{The fluctuating velocity field in a slice in the $yz$-plane at $x=0$ at different times. Downstream velocity is colour coded, in-plane velocity is shown by arrows. The length of the arrows is proportional to the in-plane component of the velocity vector. The box has been cut off at $y=2\pi$.
     The times correspond to the dots in figure~\ref{fig:re400:edgestate}. }
     \label{fig:re400:yz_slices}
    \end{figure}

    In order to study the dynamics of the edge state further, we present it in different representations at times corresponding to the circles in figure~\ref{fig:re400:edgestate}.
%     The plots in figure \ref{fig:re400:uisocontours} show isocontours of the downstream velocity component of the deviation in a slice in the $xz$-plane. Positive values are drawn as lines and negative velocities are dashed. 
    Figure \ref{fig:re400:yz_slices} shows snapshots of the fluctuating velocity field in a slice in the $yz$-plane. 
    The plots are cut off at $y=2\pi$ since in the upper part of the box no deviations from the laminar flow are visible.
    It becomes clear that the regions of high- and low-speed fluid already visible in the $x$-averaged plots are in fact streamwise high- and low-speed streaks.
    The snapshots hint at some internal dynamics, which below will be associated with a travelling-wave-like behaviour.

\begin{figure}
        \includegraphics{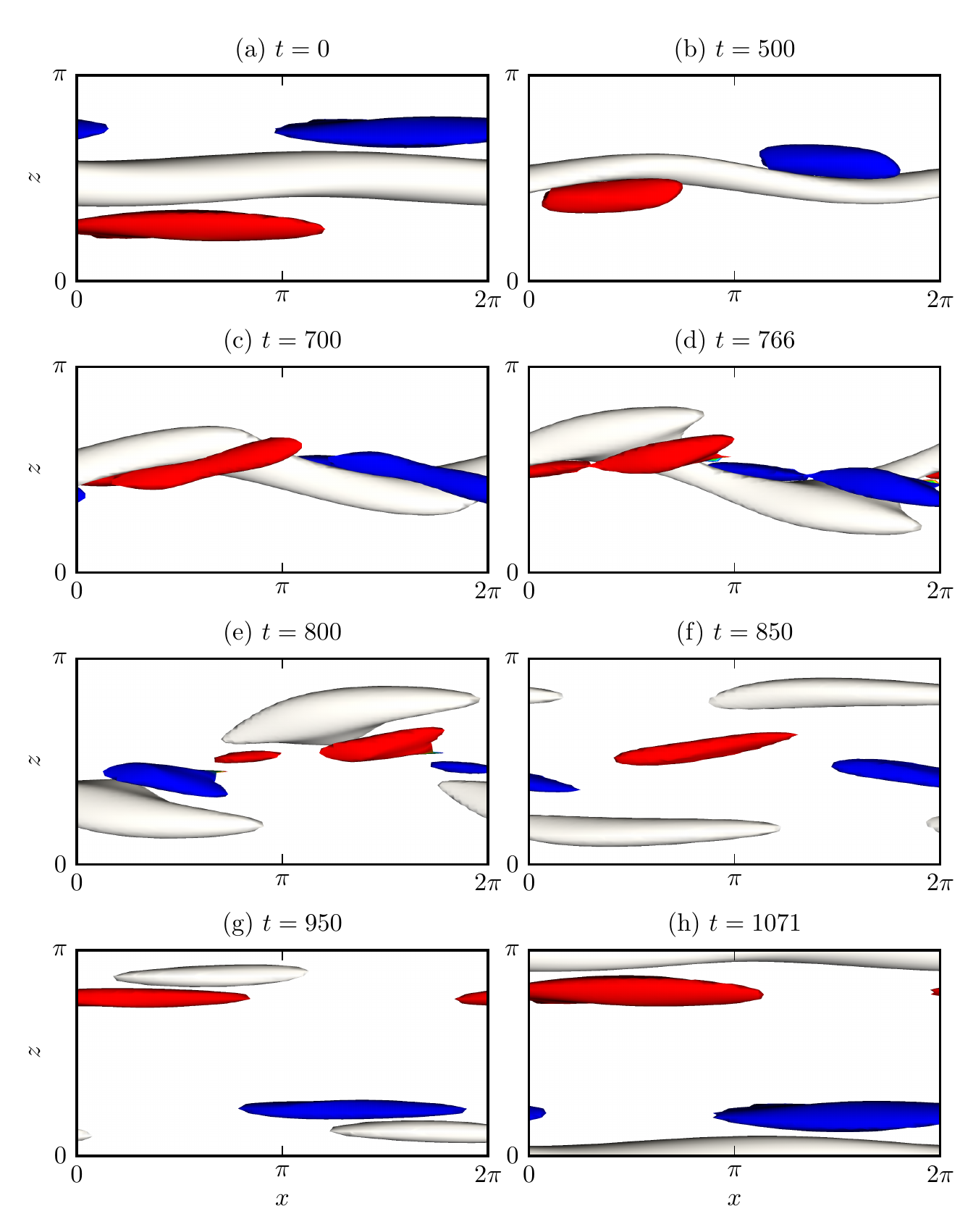}
        \caption{Snapshots of vortex-streak-interactions in a view from the top. We show the isocontour $u=-0.33$ of the deviation from the laminar profile in light gray together with isocontours of $\lambda_2 = -0.6$ \citep{Jeong1995}, where we have normalized $\lambda_2$ such that $min(\lambda_2)=-1$. The $\lambda_2$ isocontours are coloured red and blue for positive and negative downstream vorticity, respectively, with red indicating a vortex that rotates clockwise. Between $t=766$ and $t=800$ the vortices switch places, thereafter new streaks are created.}
        \label{fig:re400:lambda2}
    \end{figure}

    The in-plane components in figure~\ref{fig:re400:xaverage} suggest that vortices play an important role in the edge state's dynamics; we visualize them using the $\lambda_2$ vortex detection criterion \citep{Jeong1995}. 
    Figure \ref{fig:re400:lambda2} shows the position of vortices by isocontours of $\lambda_2$ and their interaction with the low-speed streak (isocontour $\uflucx = -0.33$, light gray) in a parallel projection, seen from the top of the box. 
    Since the strength of the vortices varies considerably, we have normalized $\lambda_2$ at every time slice such that $min(\lambda_2) = -1$ and plot the isocontour $\lambda_2=-0.6$. 
    It is coloured according to the sign of the downstream vorticity $\omega_x$ in order to distinguish vortices that rotate left and right.
    
    We start describing the sequence of events at time $t=0$ where $\ecf$ is minimal.
    The flow consists of two streamwise streaks: one low-speed streak that reaches out to approximately $\delta_{99\%}$ into the flow, and beneath it one smaller high-speed streak closer to the wall. 
    They are mostly aligned in the downstream direction, but the downstream modulation in figure~\ref{fig:re400:lambda2} shows that it corresponds to a travelling-wave-like structure. 
    The two streaks are accompanied by two streamwise elongated counter-rotating vortices, located at both sides of the low-speed streak and at different streamwise positions, just like the typical flow structures identified in \citet{Jeong1997}.
    The vortices create and sustain the streaks by linear advection of the laminar base flow. On one side a vortex lifts low-speed fluid from near the wall upwards and supports the low-speed streak. On the other side, fast moving fluid from the upper flow regions is pushed towards the wall where the base flow is slower, thereby creating a streak of high-speed fluid. Since the two vortices are counter-rotating, they both act to feed both streaks.
    For a long time the streaks persist without much change in shape while the energy in the in-plane velocity components $\ecf$ remains small. The vortices slowly grow in strength and start to ``lean'' over the low-speed streak, which in turn becomes wavier and the oscillations increase considerably.
    Approximately $700$ time units after the burst the vortices lie on top of the now very contorted low-speed streak.
    A little later the oscillations become too strong and the streaks break up -- the event corresponding to the strong burst in $\ecf$.
    Figure~\ref{fig:re400:lambda2}(d) suggests that the vortices actually switch places.
    The low-speed streak does not vanish in the process, but rather is torn apart into two smaller streaks, which then reunite across the periodic boundary conditions.
    After the break-up, the pairs of streaks and vortices are located at a spanwise location that is shifted by exactly $L_z/2$ or half a box width.
    Comparing figures~\ref{fig:re400:lambda2}(a) and (h) one finds that there is no shift in the downstream position of the structures.

\begin{figure}
        \includegraphics{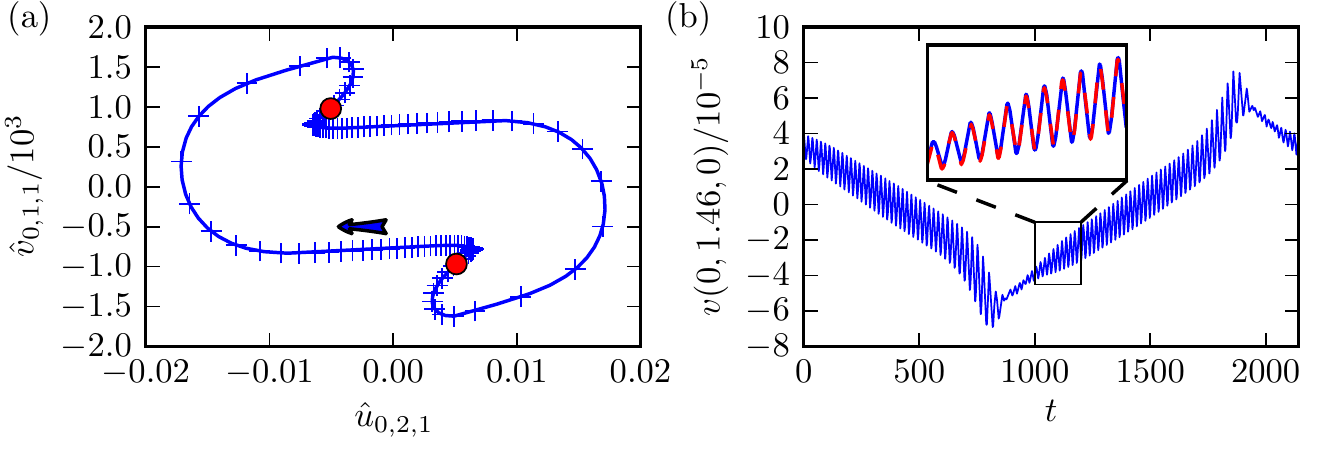}
      \caption{Time traces of the edge state. (a) A plot of the edge state trajectory in a projection of state space. The circles correspond to the minima of $\ecf$, crosses are drawn every 20 time units. The trajectory is a closed orbit passing through two minima. After a minimum of $\ecf$ the dynamics is rather slow, followed by a rapid motion as the next minimum is approached. (b) Time evolution of $\uflucy$ at $(0, 1.46,0)$. The curve shows that two timescales are involved in the process. The inlet shows the same velocity component at time $t$ and $t+2142$ and reveals that the orbit is a periodic orbit without shift in the downstream direction.}
      \label{fig:re400:statespace}
    \end{figure}

    Another representation of the periodic orbit is presented in figure~\ref{fig:re400:statespace}.
    The reduced state-space plot of two expansion coefficients in figure~\ref{fig:re400:statespace}(a) shows that the trajectory is a closed orbit in this representation. The circle in the figure indicates the point of lowest $\ecf$.
    The spacing between the crosses that are drawn at equidistant time intervals of 20 units shows that the trajectory moves rather slowly near the points,
    before rapidly going over to the other side in an event corresponding to the burst.
    In figure~\ref{fig:re400:statespace}(b) we show the time evolution of a single in-plane velocity component at a fixed position in the box, namely $\uflucy(0, 1.46, 0)$.
    The bursts show up as a slow drift from positive to negative values and back.
    An interesting feature are the short period oscillations.
    By averaging over $20$ maxima of $\uflucy(0, 1.46, 0)$ we find that the period is not constant during a cycle: directly after the burst, it is $18.4$, corresponding to a phase-speed of approximately $2/3\uinf$. The period increases slightly and just before the burst it is $20.6$.
    The inset in figure~\ref{fig:re400:statespace}(b) contains a magnification of the indicated black rectangle. In addition to the original curve, the dashed line shows the  component at time $t+2142$: that it lies exactly on top of the first curve demonstrates that the edge state is a periodic orbit and that there is no phase-shift -- the high frequency oscillations are perfectly aligned so that the fast and slow dynamics are phase-locked.
    The velocity field during one of the short periods is reminiscent of the travelling waves found also in other systems.
    
\begin{figure}
      \includegraphics[width=\textwidth]{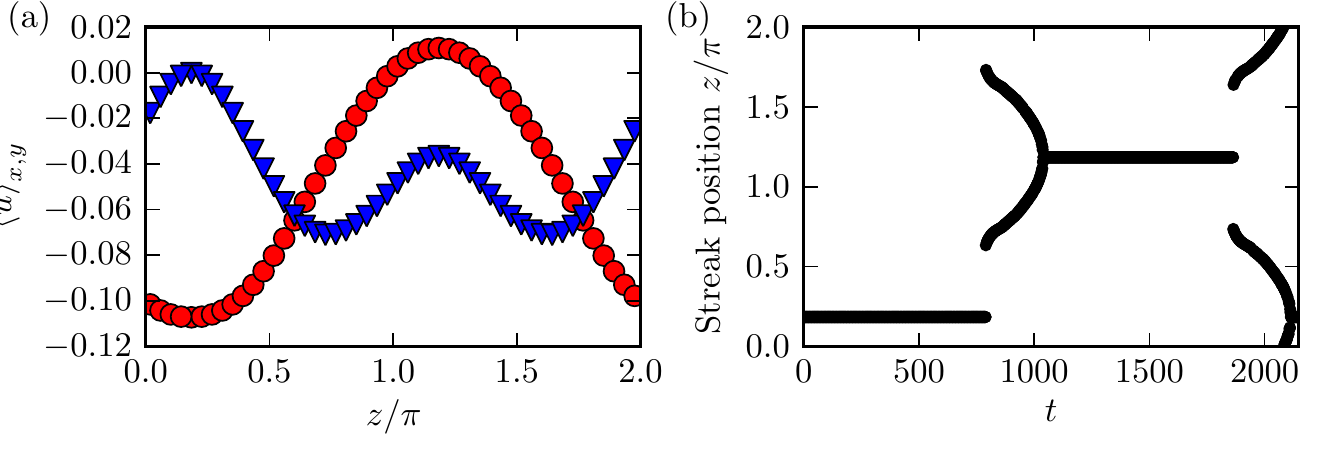}
      \caption{Localization and dynamics of low speed streaks. (a) The velocity profile of the edge state averaged over $x$ and $y$ at time $t=0$ (circles) and $t=850$ (triangles). The position of the streaks is determined by fitting a parabola through the minima of those curves. (b) The $z$-position of the low speed streaks as a function of time. During a burst, the low speed streak splits into two streaks on either side, which then reconnect to form a new one.}
      \label{fig:re400:zprof}
    \end{figure}

    A space-time plot of the streak positions shows that the shift distance is always exactly $L_z/2$.
    To determine the instantaneous position of the streaks, we average the fluctuating downstream velocity component $\uflucx$ over $x$ and $y$ and take extrema as points of reference for the location of the streak. The velocities are fitted with a parabola to locate the minima with higher precision.
    Two examples of the averaged profiles are shown in figure~\ref{fig:re400:zprof}(a); the circles show the profile at $t=0$, the triangles at $t=850$ during the burst. While in the former curve, only one low speed streak is present, the latter one shows two low speed regions.
    In figure~\ref{fig:re400:zprof}(b) we show the positions of the low-speed streaks as a function of time.
    That provides a reliable method to determine the shift distance of the structures at a burst and confirms that this distance is exactly $\pi = L_z/2$.
    Moreover, the plot shows that the streak does not shift to one side at a burst but that two streaks form to the left and right of it which then reconnect to form a single one at the new position.
    
    The edge state obeys a shift-and-reflect symmetry, that is it is unchanged by a downstream shift of $L_x/2$ combined with spanwise reflection at an appropriately chosen plane.
    Formally, it is invariant under the operation $\sigma [u,v,w](x,y,z) = [u,v,-w](x+L_z/2,y,-z)$.
    We emphasize that no symmetry has been enforced in the calculations, but that edge trajectories are drawn to this apparently attracting invariant subspace.
    It is actually this symmetry that fixes the distance of the shifts: the only other possible location for the structures in the same invariant subspace is the one related by a shift of $L_z/2$.
    This symmetry of the state also explains why the state does not drift in the spanwise direction -- a drift would require that it leaves the symmetry-invariant subspace.

  \subsection{A regenerating cycle?}
    \citet{Hamilton1995} and \citet{Waleffe1995} discuss a three step regeneration cycle for near wall turbulence.
    This quasi-cyclic process is based on the interaction between two alternating streaks and two streamwise elongated vortices.
    The streamwise vortices create the streaks by linear advection of the laminar base flow.
    But the streaks are linearly unstable to wavy modulations, that eventually lead to their break-up, causing a peak in the turbulent energy production.
    The resulting flow has a strong $x$-dependence and nonlinear interactions lead to the recreation of streamwise vortices.
    \citet{Moehlis2004} were able to capture the essential features of the process with a nine-dimensional model.
    A similar mechanism has also been used to describe a self-sustaining process for coherent structures such as travelling waves \citep{Waleffe1997}. 
    
\begin{figure}
      \centering
      \includegraphics{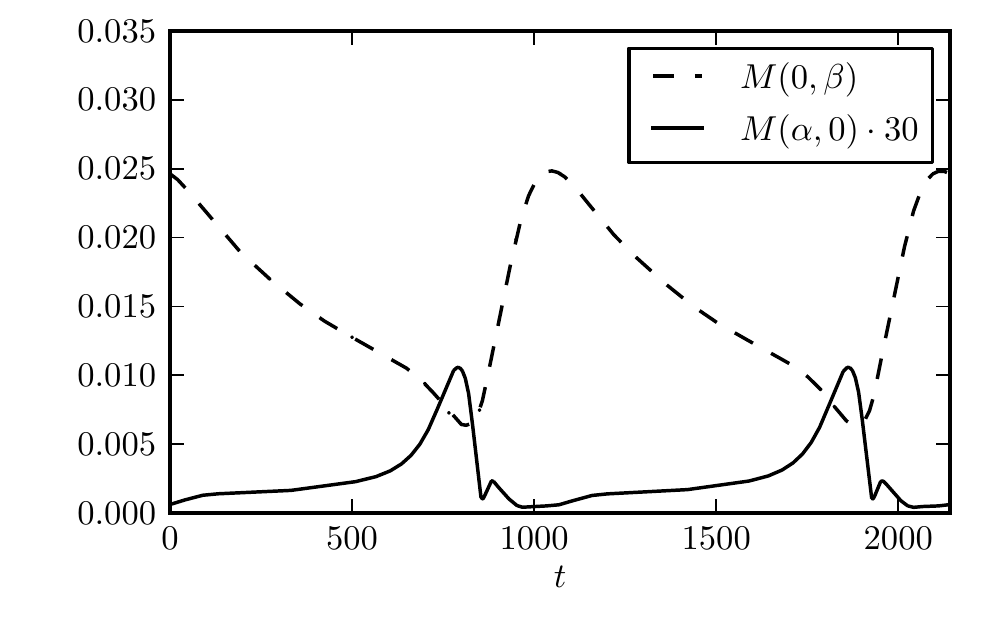}
      \caption{Modal energies $M(0,\beta)$ and $M(\alpha,0)$ as a function of time for a full cycle of the edge state. The changes in the two quantities are anticorrelated during most of the process, except for a small time interval $t=850$ to $880$. There are two phases, a long one during which $M(0,\beta)$ decreases and a short one where it increases.}
      \label{fig:re400:m0beta}
    \end{figure}

    In the edge state in the asymptotic suction boundary layer, we see structures that closely resemble the travelling waves found in pipe flow \citep{Faisst2003,Wedin2004,Hof2004} or plane Couette flow \citep{Nagata1997,Clever1997,Wang2007}. 
    They have rather high-frequency oscillations, see figure~\ref{fig:re400:statespace}(b).
    Such coherent structures have often been explained with the self-sustaining process \citep{Waleffe1997},
    but in the edge state there is a second time scale from the bursts.
    We find that the qualitative description of the low-frequency process of the bursts is more reminiscent of the process described in \citet{Hamilton1995} than the high-frequency travelling wave part.
    Important quantities for identifying the cycle are the modal energies
    \begin{equation}M(k_x=m\alpha,k_z=n\beta) = \left\{\int_{-1}^{1}\left[\hat{u}^2(m\alpha, y, n\beta) + \hat{v}^2(m\alpha, y, n\beta) + \hat{w}^2(m\alpha, y, n\beta)\right] \mathrm d y\right\}^{1/2}\end{equation}
    The general interpretation is that $M(0,\beta)$ corresponds to the streaks and $M(\alpha,0)$ to the waviness of the streaks.
    \citet{Hamilton1995} found that during the cycle $M(\alpha,0)$ and $M(0,\beta$) are strictly anticorrelated.
    In their regeneration cycle two phases of approximately the same duration are clearly distinguishable, one where the streaks grow, $\mathrm dM(0,\beta)/ \mathrm dt > 0$, and one where streaks break down, $\mathrm dM(0,\beta)/ \mathrm dt < 0$.
    We present the corresponding plot of the two quantities in figure~\ref{fig:re400:m0beta}. 
    $M(0,\beta)$ and $M(\alpha,0)$ are anticorrelated for most of the time, with the exception of a very small time interval during bursts -- possibly indicating that the mechanism that leads to the recreation of the streamwise vortices is a different one.
    The streak formation phase after a burst takes only $200$ time units, their slow breakdown lasts for more than $800$ time units.
    In the discussion of figure~\ref{fig:re400:lambda2} in the previous section, we argued that streak formation is due to linear advection by the streamwise vortices.
    A sinuous streak instability that leads to a waviness of initially straight streaks has often been observed, see for example \citet{Andersson2001} and there is little doubt that this linear instability is present in our case.
    Finally, the interactions leading to the recreation of the streamwise vortices are unquestionable nonlinear -- which means that all ingredients of the regeneration mechanism described by \citet{Hamilton1995} are present.
    
%     The question whether the process observed in the edge state in the ASBL is the same as the one responsible for the regeneration cycle described by \citet{Hamilton1995} and \citet{Moehlis2004} will be the subject of future investigations.

  \subsection{The edge state in state space}
    \label{sec:es_statespace}
    The dominance of the streaks, the long period and the dramatic bursts are reminiscent of behaviour seen in other boundary layers \citep{Kline1967}

    In a low dimensional model obtained by proper orthogonal decomposition, \citet{Aubry1988} identified an attracting heteroclinic cycle which connects two fixed points, that are related by a discrete symmetry. The fixed points have exactly one unstable eigenvalue whose real part is smaller than the modulus of the real part of the least stable eigenvalue. Trajectories spend a long time near the fixed points, before following the heteroclinic connection to the other fixed point in a wild burst.
    Motivated by these studies, \citet{Armbruster1987} undertook further analysis of the phenomenon and concluded ``that the heteroclinic cycles present in this model arise as a natural feature in the context of evolution equations which are translation and reflection invariant with respect to a spatial direction.'' \citep[p. 257]{Armbruster1987}
    Since the required $O(2)$ symmetry is also present in our system due to the periodic boundary conditions in the spanwise direction, we tried to connect our observations to their mechanisms.
    But despite considerable effort to locate the fixed points visited transiently by the trajectory, we have not been able to find them with a Newton search \citep{Viswanath2007}.
    It eventually turned out that the mechanism responsible for the bursts is different, as we now explain.
    
    The origin of the slow oscillations was found by embedding the flow in a family of flows, obtained by keeping $\uinf$ fixed and varying $\usuck$, thus exploiting the connection from ASBL to plane Couette flow (pCf).
    We work in a box of size $4\pi\times8\times2\pi$ at a resolution of $32\times65\times32$ in order to reduce the computational effort. This resolution is a little lower than for the calculations above, but the edge state dynamics is still reproduced, though the values of some variables may vary slightly; for example the bursting period T increases by less than $1\%$. Since the edge states in both the ASBL as well as in pCf have a shift-and-reflect symmetry, we enforce this symmetry for all calculations in this section.
    
%     In low Reynolds number plane Couette flow the edge state is simply an unstable fixed point \citep{Schneider2008}.
    We start in a plane Couette system with the lower plate at rest and the upper plate moving with velocity $\uinf$, where the Reynolds number based on half the velocity difference and half the gap width is $\Rey_{pCf} = \uinf H/4\nu = 800$. 
    The edge state is a fixed point in an appropriately chosen frame of reference and hence a travelling wave in the lab frame.
    The eigenvalues of the equilibrium state are calculated by means of Arnoldi iteration (implemented in channelflow by John F. Gibson, see \citet{channelflow});
    it has exactly one unstable direction, i.e.\ it is an attractor for edge trajectories.
    
\begin{figure}
      \includegraphics{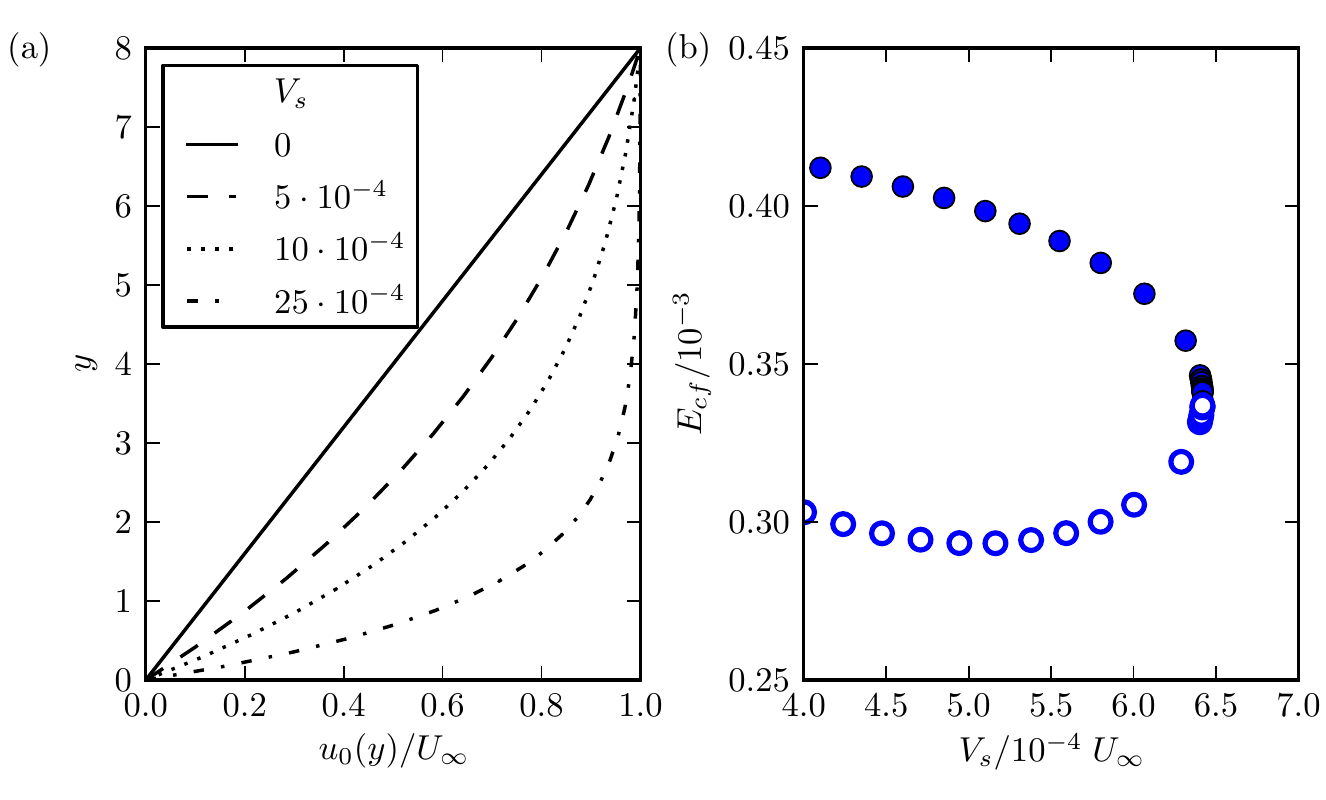}
      \caption{Connecting the asymptotic suction boundary layer to plane Couette flow.
      (a) The laminar velocity profile $u_0(y)$ for different values of $\usuck$ as indicated in the legend in a box of height $8$. For plane Couette flow ($\usuck=0$) the laminar profile is a straight line. It is smoothly transformed into the exponential profile of the ASBL as $\usuck$ increases. (b) Bifurcation diagram of the edge state (full circles). At $\usuck=6.418\cdot10^{-4}$, corresponding to $\delta=3.89$, it collides with a lower branch equilibrium (open circles) in what looks locally like a saddle-node bifurcation. Restricted to the edge, the upper branch equilibrium is stable while the lower branch has one unstable eigenvalue.}
      \label{fig:homotopy}
    \end{figure}

    The suction velocity $\usuck$ serves as a homotopy parameter between pCf and the ASBL. 
    As $\usuck$ increases from $0$, the laminar profile gradually transforms from a straight line to an exponential, figure~\ref{fig:homotopy}(a), while the boundary layer thickness $\delta=\nu/\usuck$ decreases from $\infty$ to $1$.
    Starting with the relative equilibrium that is the edge state in the plane Couette system, we continue the solution in $\usuck$;
    it vanishes in a saddle-node bifurcation at $\usuck/\uinf = 6.418\cdot10^{-4}$.
    Restricted to the edge the edge state is a node and it corresponds to the upper branch of the saddle-node bifurcation. We were able to continue it to the lower branch which has two unstable eigenvalues. 
    The bifurcation diagram $\ecf$ versus $\usuck$ is shown in figure~\ref{fig:homotopy}(b).
    Very close to the bifurcation, the leading eigenvalues, without the neutral one, of the node and the saddle are
    $0.0051$, $-3.76\cdot10^{-5}$, $-0.0022\pm0.0136i$ and $0.0053$, $2.36\cdot10^{-5}$, $-0.0026\pm0.0137i$, respectively.
    The difference occurs in the next to leading eigenvalue, which is real, indicating a standard saddle-node bifurcation.

\begin{figure}
      \centering
      \includegraphics{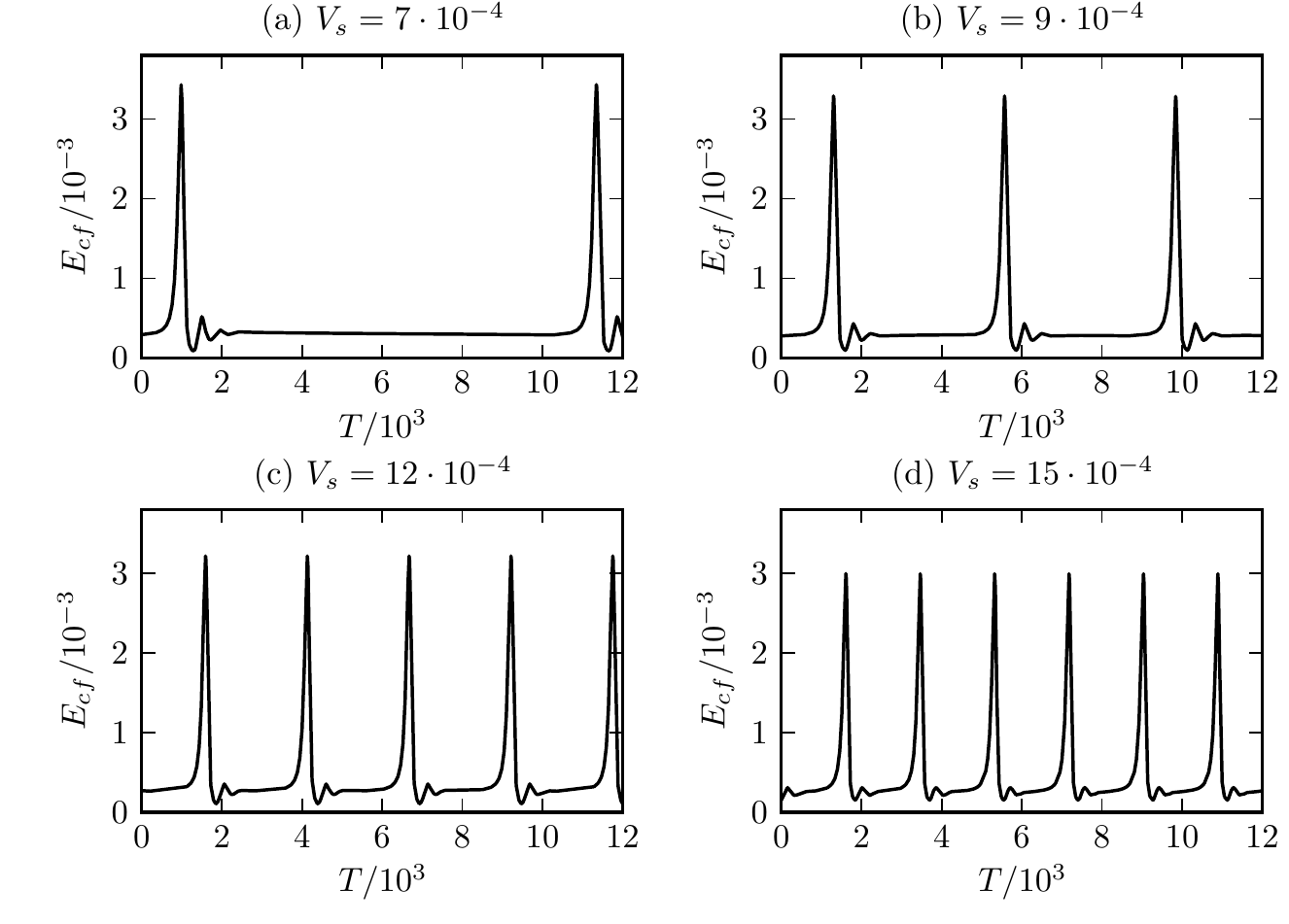}
      \caption{Cross-flow energy of edge states for several values $\usuck$. After the bifurcation point, periodic bursts set in with an extremely long period. The period decreases rapidly as the distance from the bifurcation point increases.}
      \label{fig:hom:edgetracking}
    \end{figure}

    In figure~\ref{fig:hom:edgetracking} we present the results of edge state tracking for four values of $\usuck$ after the critical value.
    Directly after the bifurcation, periodic bursts set in with an extremely long period. The period rapidly decreases with increasing distance from the critical suction value, as can be seen in figure~\ref{fig:hom:edgetracking}(b-d). At every burst, the state shifts by $L_z/2$.
  
\begin{figure}
      \centering
      \includegraphics[width=.9\textwidth]{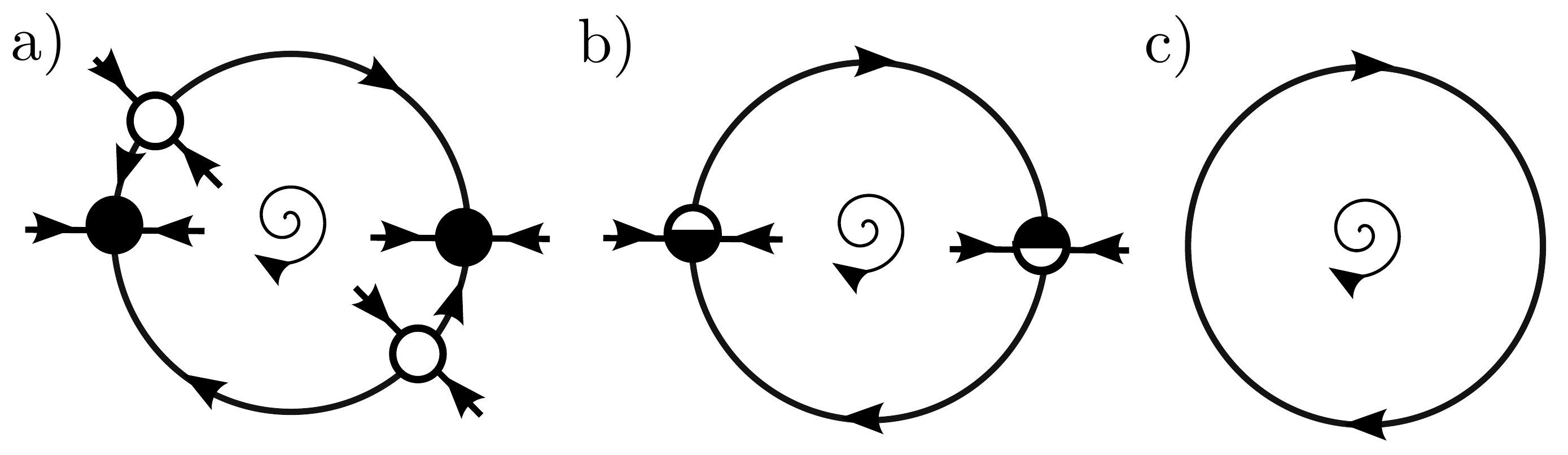}
      \caption{Schematic representation of the bifurcation in state-space. a) for small $\usuck$ two pairs of fixed points, one stable (full circles), one unstable (open circles), exist which are related by a discrete symmetry, namely shifting by $L_z/2$; they are heteroclinically connected as indicated. b) as $\usuck$ increases, the fixed points annihilate in a saddle-node infinite-period (SNIPER) bifurcation c) The fixed points are gone for larger $\usuck$, the edge state now goes around the cycle. For $\usuck$ near the bifurcation value, the period of the cycle diverges due to the ghost left behind by the saddle-node bifurcation. The stable directions shared by the fixed points and the periodic orbit contain both real and complex stable eigenvalues.}
      \label{fig:hom:snic}
    \end{figure}

\begin{figure}
      \centering
      \includegraphics{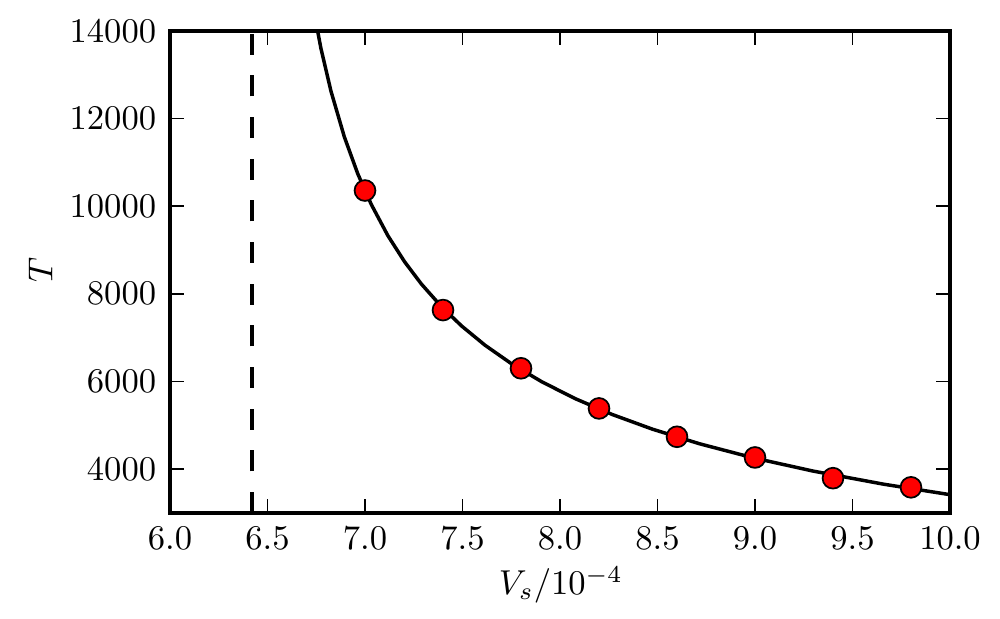}
      \caption{Divergence of the bursting period close to the bifurcation point. The fit (solid line) to the expected theoretical scaling of a SNIPER bifurcation, $T \propto 1/\sqrt{\usuck - {\usuck}_c}$,
%       , gives $T = 88.30 / \sqrt{V_s - 6.42\cdot10^{-4}} - 1249.32$. The fit 
      agrees very well with the measured data (circles). Moreover, the critical value for $\usuck$ in the fit is $6.42\cdot10^{-4}$ (indicated by a dashed line) in very good agreement with the value obtained from the saddle-node bifurcation in figure~\ref{fig:homotopy}(b).}
      \label{fig:hom:sniperscaling}
    \end{figure}

    This behavior can be explained if the edge state is heteroclinically connected with a copy of it that is obtained by shifting it by $L_z/2$ in the spanwise direction.
    Such heteroclinic connections between symmetry related fixed points have been found in several studies with $O(2)$-symmetry, see  \citet{Aubry1988, Armbruster1987, Tuckerman1988, Halcrow2008, Gibson2008}
    Since both states are attractors (in the same symmetry-subspace inside the edge),  they have to be separated by a saddle.
    For symmetry reasons, the heteroclinic connection exists as well in the other direction, making it an attracting invariant cycle.
    The resulting state-space structure is shown in figure~\ref{fig:hom:snic}(a); the unstable direction of the edge points perpendicular to the plane.
    The global bifurcation is hence a saddle-node infinite-period (SNIPER) bifurcation \citep{Strogatz1994,Tuckerman1988,Abshagen2005}, also known as saddle-node-on-invariant-cycle (SNIC).
    One indicator of a SNIPER bifurcation is a divergence of the period (figure~\ref{fig:hom:snic} b,c), which can already be read off from figure~\ref{fig:hom:edgetracking}.
    More quantitatively, one expects the period to diverge like $1/\sqrt{V_s - {\usuck}_c}$ at the bifurcation point.
    The data shown in figure~\ref{fig:hom:sniperscaling} together with the fit,
    $T = 88.30 / \sqrt{V_s - 6.42\cdot10^{-4}} - 1249$, 
    is in very good agreement with this expectation.
    Also the critical value for $\usuck$ obtained from the fit is $6.42\cdot10^{-4}$, in excellent agreement with the value obtained from the saddle-node bifurcation in figure~\ref{fig:homotopy}(b).

\begin{figure}
      \centering
      \includegraphics{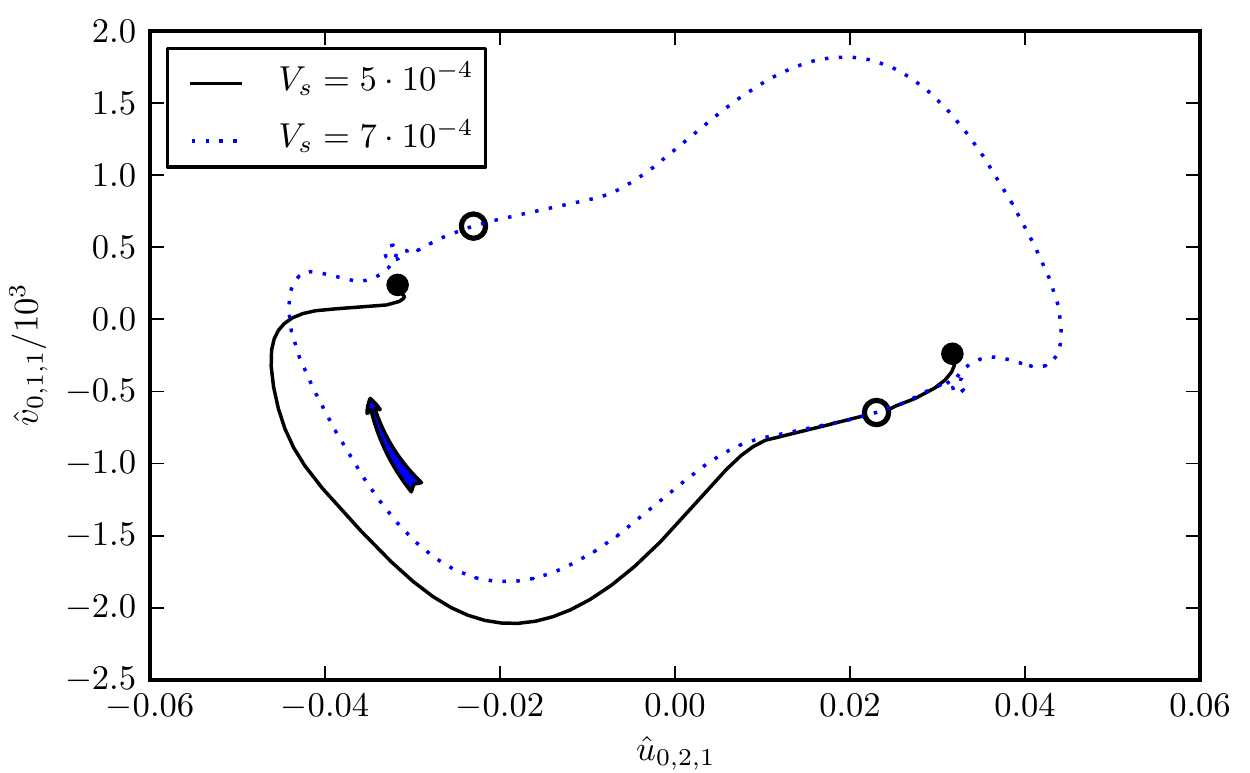}
      \caption{The heteroclinic connection in state-space at $\usuck = 5\cdot10^{-4}$ in a two dimensional projection. We show the node (full circle) and the saddle (open circle) together with the pair of fixed points obtained by a shift of $L_z/2$ (the projection is independent of $x$). The black lines are obtained by edge tracking from initial conditions which are slight perturbations of the saddle point; they reveal the heteroclinic connections from the saddle to both stable fixed points. The dotted line is the (bursting) edge state at $\usuck=7\cdot10^{-4}$, slightly above the bifurcation.}
      \label{fig:hom:hetcon}
    \end{figure}

    Further evidence for this scenario comes from a study of the explicit identification of the heteroclinic connection.
    The procedure we used to identify the connection is as follows:
    restricted to the edge of chaos, the node is an attractor while the saddle has one unstable direction.
    According to figure~\ref{fig:hom:snic}, an edge-trajectory that starts near the saddle will follow the heteroclinic connection either to the node or the shifted node.
    Using the edge state tracking algorithm to restrict the trajectories to the edge, we start from initial conditions near the saddle.
    A projection of state-space in figure~\ref{fig:hom:hetcon} shows the two pairs of node and saddle, the heteroclinic connection to both sides as well as an edge state after the bifurcation.
    
    The long phase between two bursts of the edge state in the ASBL does hence not correspond to the transient visit of a weakly unstable fixed point but rather to the ghost left behind by a saddle-node bifurcation on an invariant cycle. 
    
\section{Variation of flow parameters}
\label{sec:variation}
  We now turn to a study of the properties of the edge state at different Reynolds numbers and in different domains.
  Roughly $200$ edge states have been calculated for different parameter values.  
  While there are some changes in detail, the edge states' structure is preserved in all cases. It consists of two streamwise streaks and two streamwise elongated counter-rotating vortices and there are periodic bursts where the state is shifted by half a box width.

  The criterion for the convergence of a calculation is based on the cross-flow energy. We only consider states where the evolution of $\ecf$ settles to a clearly identifiable asymptotic state, either constant or periodic.
  We found that all states falling into this category display periodic bursts.
  Convergence is assumed if $\ecf$ has at least three peaks where their height and the time between the peaks agree within a relative tolerance of $2\%$.

  \subsection{Reynolds number}
    \label{sec:reynolds}

    In this section the Reynolds numbers is varied while the box size is kept fixed at $4\pi\times2\pi$. Calculations still take place in the transitional regime, with Reynolds numbers between $320$ and $750$. 
    Below $\Rey = 320$ the cross-flow energy at bursts is of the same order as the cross-flow energy of fully turbulent states and the distinction between the edge state and turbulent motion becomes difficult as in pipe flow \citep{Schneider2009}. 
    The highest Reynolds number we calculated is $750$. There is, however, no indication that the structure of edge states would change for slightly higher Reynolds numbers.

\begin{figure}
      \centering
      \includegraphics{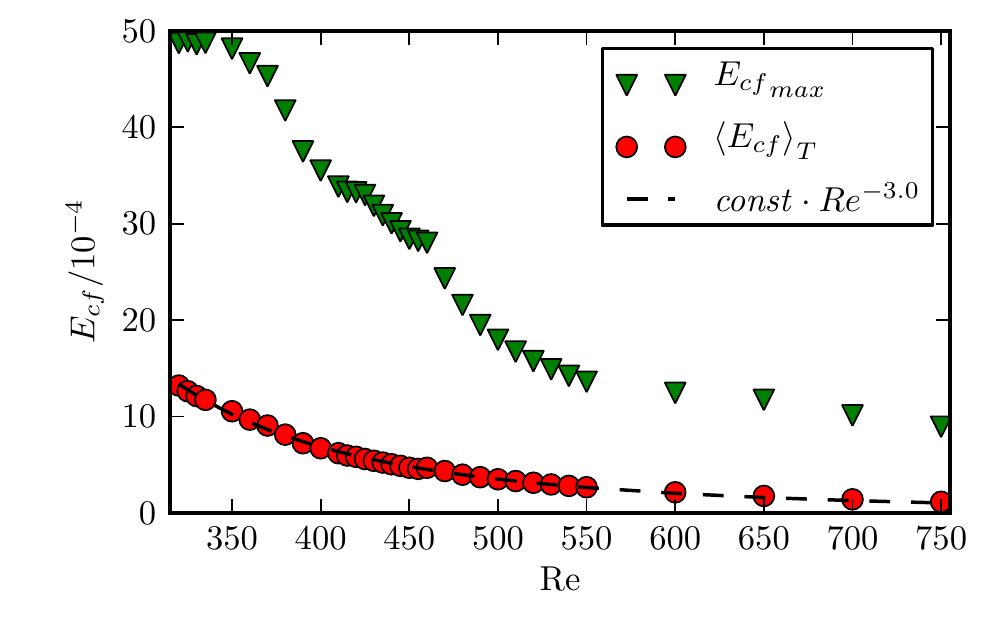}
      \caption[Cross-flow energy vs. $\Rey$]{The energy of the cross-flow $\ecf$ as a function of $\Rey$. Green triangles: maximum of $\ecf$ at bursts; red circles: $\ecf$ averaged over one period.
      The dashed line is a fit to a power law, $\ecf\sim \Rey^{-3}$.
      The energy generally decreases with increasing $\Rey$.}
      \label{fig:re:ecf}
    \end{figure}

    Figure~\ref{fig:re:ecf} shows the cross-flow energy $\ecf$ as a function of Reynolds number. The maximum value of $\ecf$ at a burst is plotted with triangles. This value may be taken as a measure for the burst intensity, 
    it decreases monotonically with Reynolds number.
    A potential explanation comes from the observation, that at higher Reynolds numbers the critical amplitude for triggering transition to turbulence is lower \citep{Levin2005} and that the edge state's energy reflects this descending threshold.
%     . An interpretation for this is right at hand: 
%     it is well known that for increasing Reynolds numbers the critical amplitude for triggering transition to turbulence decreases \citep[e.g.][]{Levin2005}.
%     Since the states selected by the edge state tracking algorithm are on the border of the chaotic saddle that is turbulence, their energy content reflects this descending threshold.
    The circles show the time averaged cross-flow energy.
    It decreases monotonically and smoothly with $\Rey$ for the same reasons as the maximum energy.
%     An exponential dependence can be safely excluded based on the available data. % a power law seems to fit better.
    % The range of Reynolds numbers is too small to give a reliable estimate of the exact functional dependence.
    A power law gives the best fit with ${\langle\ecf\rangle}_T \sim \Rey^{-3}$, plotted as a dashed line in the same figure.

  \subsection{Streamwise and spanwise domain-size}
    \label{sec:boxsize}
%     \subsection{General results}
    In this section, we study edge states at Reynolds number $400$ but in boxes of different size. $L_x$ varies from $3.2\pi$ to $7\pi$ in steps of $0.2\pi$, with $L_z$ ranging from $L_x/2 - 0.5\pi$ to $L_x/2 + 0.4\pi$ in steps of $0.1\pi$.
    In smaller boxes it was difficult to find sustained turbulent states.

\begin{figure}
      \centering
      \includegraphics{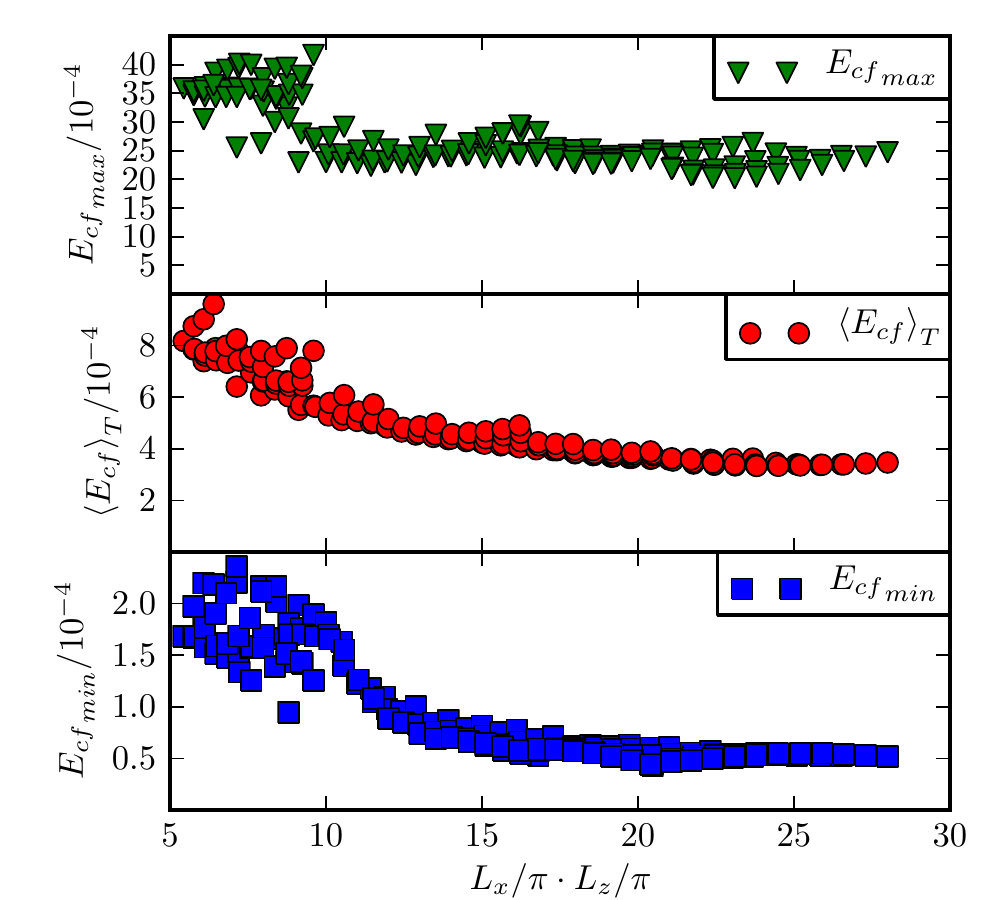}
      \caption{The cross flow energy $\ecf$ as a function of the boundary layer volume. Two qualitatively different regions can be distinguished: states with $L_x\cdot L_z < 10$ are somewhat higher in energy and show wide variations. For larger values the maximum of $\ecf$ seems to prefer a value near $25$ and the variations in all energies are reduced.
%       The third group of states, marked with open symbols, are from the boxes around $5.4\pi\times2.5\pi$. They lie outside the general trend and seem to form a group of their own. Theses states include, among others, the period doubled state shown in figure~\ref{fig:perioddoubling}.
      }
      \label{fig:bs:lxlzecf}
    \end{figure}

    In figure~\ref{fig:bs:lxlzecf} the cross-flow energy is plotted as a function of the boundary layer volume $L_x\cdot L_z$. 
    We distinguish two qualitatively different regions: small boxes with $L_x L_z \lesssim 10\pi^2$ and larger boxes with $L_x L_z \gtrsim 10\pi^2$.
    In larger boxes, the variation of the cross-flow energy is rather smooth. $\ecfmax$ is roughly constant in those boxes: the burst intensity does not depend on the box size if the box is large enough. 
    Both $\ecfmin$ and $\ecfav$ decrease for increasing box size, but the variation is very small for the largest boxes.
    In the small boxes, the minimum and time-averaged cross-flow energy are not constant as a function of box size.
    States in these boxes seem to depend heavily on the available space and the greatest possible wavelength. 
    Many states in these boxes have a comparable value of $\ecfmax$ which is roughly $50\%$ higher than in larger boxes.

\begin{figure}
      \centering
      \includegraphics{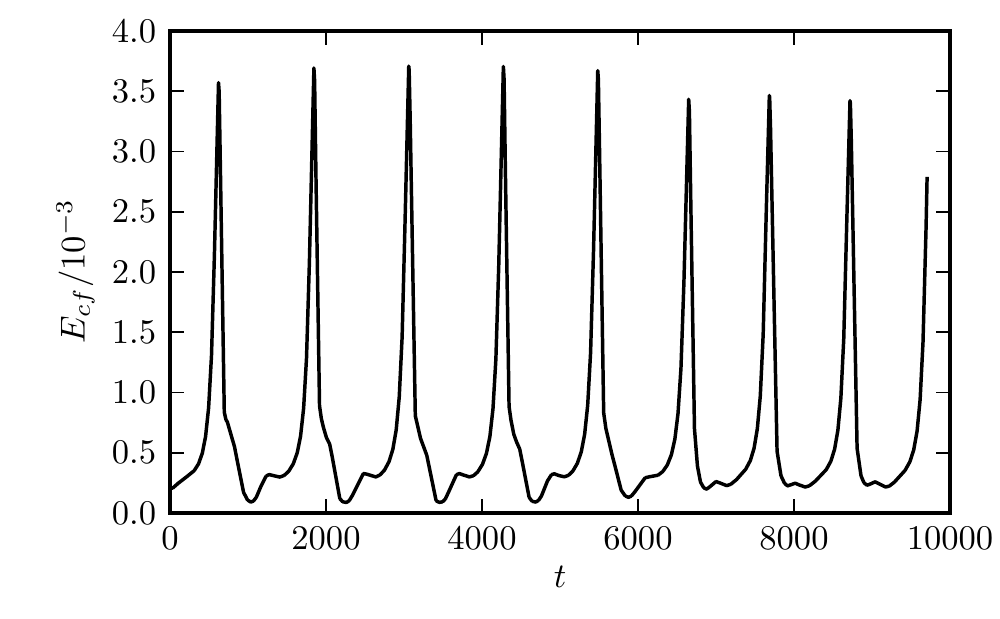}
      \caption{Non-converged edgetracking evidenced by the cross-flow energy in a box of size $4.2\pi\times2\pi$. 
      Up to a time of about $6000$ the tracking seems to converge rapidly. However, beyond that the curve changes: the bursts are less intense, the bursting period is a little lower -- the edge state tracking has converged to another state, even though the first one would have been considered as converged by the criterion described at the beginning of this section.}
      \label{fig:shapechange}
    \end{figure}

% 
%     The states from boxes around $5.4\pi\times2.5\pi$, marked with open symbols, lie outside the rather smooth variation of the curves, both $\ecfmax$ and $\ecfav$ are considerably higher and also $\ecfmin$ varies more. 
%     The burst intensity, $\ecfmax$, in these boxes has approximately the same value as in smaller boxes.
    There are some states, mainly from boxes around $5.4\pi\times2.5\pi$, where the edge trajectory seems to converge to one state, before switching over to another one, with less amplitude in the bursts.
    As an example, in figure~\ref{fig:shapechange} we present an edge state tracking in the box $4.2\pi\times2\pi$ where that happens.
    We first see five peaks of the same height and at equal distance, where the edge state tracking seems to be converged. But then the curve changes and another state is approached, where $\ecfmax$ and $\ecfmin$ are lower and higher, respectively.
    We explain this behavior as follows: there exist two periodic orbits in these boxes, one of which is stable inside the edge and the other one is weakly unstable. 
    The edge trajectory approaches the unstable one first and lingers so close to it that it seems to be converged.
    But if the integration is carried on long enough, it leaves this unstable state and finally converges to the edge state.
%     As an example, in figure~\ref{fig:shapechange} we present an edge state tracking in the box $4.2\pi\times2\pi$ where that happens.
    Figure~\ref{fig:shapechange} shows that sometimes edge tracking has to be continued for very long times before converged results are obtained. It would be possible to distinguish transient states from true periodic states if they could be identified numerically exactly, say using Newtons method. Then the differences in the number of unstable eigenvalues would show up. However, it has not been possible to achieve this because of the long periods.
    Given the large number of calculated states we feel that the overall trends discussed in this section are reliable.
    
    \subsection{Burst period}
    The trend in the bursting period $T$ is inverse to that in cross-flow energy. 
    The period is small for low Reynolds numbers and grows with increasing $Re$. 
    Similarly, it is small for small boxes and larger in larger boxes.

  \subsection{A breaking of the space-symmetry}
    \label{sec:bs_perioddoubling}

\begin{figure}
      \centering
      \includegraphics{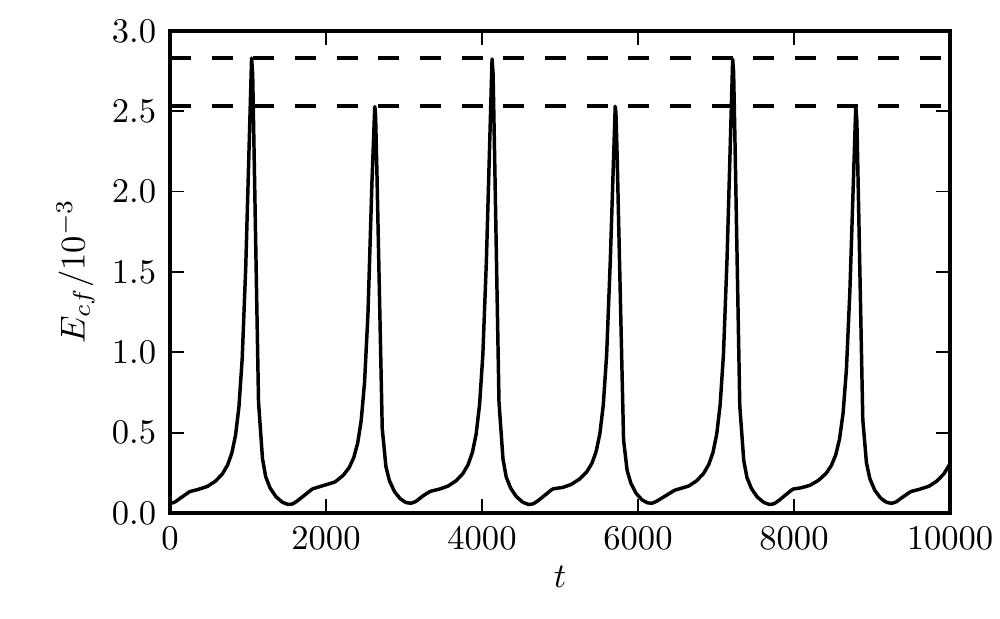}
      \caption{Cross-flow energy of the edge state in the box $5.6\pi\times3\pi$. The burst intensity is no longer the same for every burst, but alternates between a lower and a higher value. The dashed lines correspond to the two maxima and confirm the periodicity of the state.}
      \label{fig:perioddoubling}
    \end{figure}

    Figure~\ref{fig:perioddoubling} shows the cross-flow energy of the edge state in the box $5.6\pi\times3\pi$. The burst intensity, measured by the maximum of $\ecf$, is no longer the same for every burst, it alternates between a lower and a higher value. The periodicity of $\ecf$ is not completely lost, but the time interval after which it repeats has doubled. 
    Still, at every burst the position of the streaks is shifted by half a box width, but the symmetry between the two positions is broken.
%     The period of the periodic orbit includes two bursts, due to the symmetry shifts. It does hence not change with the period doubling we see here.
%     The period of the periodic orbit that is the edge state is not affected since it is twice the bursting period. The edge states' structure is still preserved, at a burst the streaks are shifted by exactly $L_z/2$. 
    
\section{Conclusions}
  \label{sec:conclusions}
%  We fully characterize a periodic orbit which acts as the edge state in the asymptotic suction boundary layer. To our knowledge this is the first known exact invariant solution in a boundary layer geometry.
  
  The periodic orbit in the asymptotic suction boundary layer which we have characterized here 
  shows dynamics on two different time scales:
  similar to invariant states in plane Couette and pipe flow, it shows the typical advection of downstream vortices on a fast time scale.
  Regular energetic burst events, not unlike those that have been reported as a key feature of boundary layer flows, occur on a much slower time scale.
  Although the flow can be turbulent and the dynamics is characterized by positive Lyapunov exponents, the two time scales remain phase-locked over the whole period of several thousand advective time units. In view of the translational symmetry of the
  asymptotic suction boundary layer these phenomena could well be the periodically continued version of the
  evolving boundary layer edge state described by \citet{Duguet2012}.

  For the bursting dynamics, an explanation from two different view points is given.
  Physically, the bursting is the result of vortex-streak interactions.
  A pair of downstream vortices creates and sustains the streaks by linear advection.
  The low-speed streaks are linearly unstable and develop a waviness, which in turn causes the vortices to become tilted with respect to the $x$-axis.
  The two counter-rotating vortices ``lean'' over the low-speed streak, tearing it apart, and eventually switch sides.
  Afterwards, the streaks are recreated by the vortices and the process begins anew, only shifted by half a box width.
  From a dynamical systems point of view, we are able to associate the bursting with a SNIPER-bifurcation:
  in plane Couette flow, two symmetry-related pairs of saddle and node are heteroclinically connected. As the homotopy parameter,
  the suction velocity, is increased, each of the pairs collides in a local saddle-node bifurcation, leaving behind a periodic orbit along the former heteroclinic connections.
  
  The orbits discussed here are linearly unstable. They have a single unstable direction and are part of the edge that controls the transition to turbulence.
  Typically, they are lower branch states in a saddle-node bifurcation where the upper branch state becomes part of the turbulent dynamics.
  One can then expect, though this deserves further study, that the upper branch state shows a similar dynamics.
  Solutions like the one discussed here, even when they are unstable, form a backbone for the turbulent dynamics, as they are transiently visited by a turbulent trajectory.
  Then the long periodic dynamics provides a mechanism for energetic bursts and adds one more possibility to the list of bursting phenomena discussed by Robinson (1991). 

  The obvious spatial localization in the edge calculations in the Blasius boundary layer \citep{Cherubini2011,Duguet2012}, as well as similar structures in confined flows, suggests the existence of spatially localized counterparts of the periodic orbit discussed here, as indeed found for spanwise localization in \citet{Khapko2013}.
  We expect that further comparisons between these and other features found in the asymptotic suction boundary layer and the full Blasius boundary layer will help to identify and characterize shared generic features and to separate them from suction related
  influences.
  
%   There is some controversy regarding the role of horseshoe or hairpin vortices in the dynamics of turbulence production associated with bursts \citep{Robinson1991}.
%   We just note that that the bursts in the edge state don't invole such structures.
%   The question whether they are important for the transition from the edge state to turbulent motion has not been part of this study.

\subsection*{Acknowledgements}
  We thank John F Gibson for providing and maintaining the open source Channelflow.org code.
  We thank Taras Khapko, Yohann Duguet, Philipp Schlatter and Dan S Henningson for discussions and the exchange of results and Stefan Zammert and Fernando Mellibovsky for discussions.
  
% \bibliographystyle{jfm}
% \bibliography{library}
% \bibliography{library}
% \bibliography{references}

\end{document}